%% file: aaai25.tex
\documentclass[letterpaper]{article} % DO NOT CHANGE THIS
\usepackage{aaai25}  % DO NOT CHANGE THIS
\usepackage{times}  % DO NOT CHANGE THIS
\usepackage{helvet}  % DO NOT CHANGE THIS
\usepackage{courier}  % DO NOT CHANGE THIS
\usepackage[hyphens]{url}  % DO NOT CHANGE THIS
\usepackage{graphicx} % DO NOT CHANGE THIS
\usepackage{natbib}  % DO NOT CHANGE THIS AND DO NOT ADD ANY OPTIONS TO IT
\usepackage{caption} % DO NOT CHANGE THIS AND DO NOT ADD ANY OPTIONS TO IT
\usepackage{multirow}
\usepackage{booktabs}
\usepackage{amsmath} 
\usepackage{xspace}
\usepackage{algorithm}
\usepackage{algorithmic}

\newcommand{\DMF}{\text{Data and Model Factory}\xspace}
\newcommand{\PR}{\text{PR Insights}\xspace}
\newcommand{\InstructLab}{\text{name anonymized}\xspace} 
\usepackage{newfloat}
\usepackage{listings}
\DeclareCaptionStyle{ruled}{labelfont=normalfont,labelsep=colon,strut=off} % DO NOT CHANGE THIS
\floatstyle{ruled}
\newfloat{listing}{tb}{lst}{}
\floatname{listing}{Listing}
\title{Deploying Privacy Guardrails for LLMs: A Comparative Analysis of Real-World Applications}

\author{
    Shubhi Asthana, 
    Bing Zhang,
    Ruchi Mahindru, 
    Chad DeLuca,  \\
    Anna Lisa Gentile,
    Sandeep Gopisetty
}
\affiliations{
    IBM Research\\
    sasthan@us.ibm.com,
    bing.zhang@ibm.com,
    rmahindr@us.ibm.com,
    delucac@us.ibm.com, \\ 
    annalisa.gentile@ibm.com,
    sandeep.gopisetty@us.ibm.com
}

\usepackage{bibentry}
\begin{document}

\maketitle

\begin{abstract}
The adoption of Large Language Models (LLMs) has revolutionized AI applications but poses significant challenges in safeguarding user privacy. Ensuring compliance with privacy regulations such as GDPR and CCPA while addressing nuanced privacy risks requires robust and scalable frameworks. This paper presents a detailed study of OneShield Privacy Guard, a framework designed to mitigate privacy risks in user inputs and LLM outputs across enterprise and open-source settings. We analyze two real-world deployments: (1) a multilingual privacy-preserving system integrated with \DMF, focusing on enterprise-scale data governance; and (2) \PR, an open-source repository emphasizing automated triaging and community-driven refinements.

%Through a comparative evaluation, we highlight the strengths, limitations, and lessons learned from each deployment. The first deployment achieved superior multilingual detection accuracy with regulatory adaptability, while the second deployment demonstrated the value of lightweight solutions and human-in-the-loop mechanisms for iterative improvement. Insights from these deployments reveal critical considerations for context-aware entity recognition, policy enforcement, and ethical AI adoption. This work contributes to advancing privacy-preserving AI systems, supporting user trust and compliance in diverse operational contexts.
In Deployment 1, OneShield achieved a 95\% F1 score in detecting sensitive entities like dates, names, and phone numbers across 26 languages, outperforming state-of-the-art tools such as StarPII and Presidio by up to 12\%. Deployment 2, with an average F1 score of 0.86, reduced manual effort by over 300 hours in three months, accurately flagging 8.25\% of 1,256 pull requests for privacy risks with enhanced context sensitivity.

These results demonstrate OneShield's adaptability and efficacy in diverse environments, offering actionable insights for context-aware entity recognition, automated compliance, and ethical AI adoption. This work advances privacy-preserving frameworks, supporting user trust and compliance across operational contexts.
\end{abstract}


\input{introduction} 
\input{state_of_art}
\input{deployment1}

\input{deployment2}

\input{comparison}

\input{conclusion}

%\subsubsection{References.}

\bibliography{aaai25}

\end{document}

%% file: introduction.tex
\section{Introduction}
\label{sec:Introduction}

As Large Language Models (LLMs) gain widespread adoption in consumer and enterprise applications, ensuring data privacy has become a critical challenge. These models often process vast amounts of unstructured data that may contain sensitive Personally Identifiable Information (PII). For example, the latest Common Crawl dataset \cite{smith2013dirt} encompasses data from over 3 billion web pages and has vast user PII data. This raises concerns about data leakage, compliance with global regulations, and ethical use \cite{carlini2021extracting, carlini2022quantifying}. While LLMs enable innovative applications, their ability to unintentionally store and recall sensitive data during inference creates vulnerabilities that must be addressed through robust privacy-preserving mechanisms.

LLMs process unstructured data, ranging from text prompts to conversational outputs, often contains sensitive details like names, phone numbers, and credit card information. Furthermore, data sourced globally may inadvertently violate region-specific privacy regulations, including the General Data Protection Regulation (GDPR) \cite{gdpr}, California Consumer Privacy Act (CCPA) \cite{ccpa}, and Personal Information Protection and Electronic Documents Act (PIPEDA) \cite{pipeda}. The scale and complexity of LLM deployments amplify compliance challenges.

Moreover, traditional privacy techniques, such as rule-based regular expression (regex) methods or standalone machine learning models, often struggle to balance precision, contextual understanding, and computational efficiency. Detecting nuanced privacy risks, such as context-dependent sensitivity, requires advanced mechanisms. Simple entity recognition is not enough. For example, consider an office phone number publicly listed on a company website for media inquiries. This is different from the personal phone number of a private individual, like John Doe. Handling these cases requires careful analysis to comply with privacy regulations and prevent unintended disclosure. Consequently, deploying privacy-preserving systems for LLMs demands innovative framework that address these multifaceted challenges while ensuring scalability and real-time responsiveness.

To address these questions, we designed a framework \textbf{OneShield Privacy Guard}. The key contributions of the different stages in the framework are as follows:
\begin{itemize}
\item A detailed analysis of two distinct OneShield Privacy Guard deployments addressing different application scenarios.
\item A comparison of their technical architectures, privacy-preservation methods, and effectiveness.
\item Insights into scalability, compliance, and lessons learned that inform best practices for future LLM deployments.
\end{itemize}

This work advances the understanding of deploying privacy safeguards in diverse LLM environments, contributing to the field of privacy-preserving AI and supporting the broader ethical use of LLM technologies.

This paper focuses on two deployments of the OneShield Privacy Guard framework, examining their distinct approaches to addressing privacy risks in real-world environments. By analyzing these cases, the paper aims to shed light on best practices and lessons learned from deploying privacy-preserving solutions in diverse operational contexts. Specifically, the objectives include:
\begin{enumerate}
    \item \textbf{Exploring Deployment Scenarios:}
    \begin{enumerate}
        \item \textbf{Deployment 1:} Safeguarding LLM outputs in the  Data and Model Factory platform, which is an internal platform designed to provide AI models and datasets with robust data governance capabilities. The safeguard focuses on enterprise-scale data governance and multilingual privacy compliance.
        \item \textbf{Deployment 2:} Securing community-contributed datasets in the open-source repository \textit{\InstructLab}, ensuring adherence to project codes of conduct and automated privacy triaging.
    \end{enumerate}
    \item \textbf{Comparative Analysis of Deployments:}
        \begin{enumerate}
        \item Identify the similarities and differences in their architectures, privacy-preservation methods, and performance metrics.
        \item Highlight the strengths and limitations of each deployment in addressing contextual privacy concerns.
        \end{enumerate}
    \item \textbf{Scalability and Adaptability:}
      \begin{enumerate}
        \item Assess the scalability of the OneShield Privacy Guard framework across languages, domains, and operational environments.
        \item Discuss the adaptability of these techniques to evolving privacy standards and emerging data protection laws.
    \end{enumerate}
\end{enumerate}

In the remainder of this paper, we first review the existing state-of-the-art PII works, followed by descriptions of our two deployments. We then perform a comparison study between the two deployments and discuss our lessons learned from them. Finally, we conclude the paper and discuss future work.

%% file: state_of_art.tex
\section{State of the Art}
\label{sec:state-of-art}

\textbf{Existing Techniques for PII Detection in LLMs:} Detecting Personally Identifiable Information (PII) within Large Language Models (LLMs) is an area of growing interest, with research focusing on developing scalable and accurate solutions. Traditional methods rely on regular expression (regex)-based tools like \cite{comprehend} and \cite{neuralseek}, and customized machine learning models to identify PII entities like names, addresses, date, social security numbers, Vehicle Identification Number (VIN), bank account numbers etc. For instance, tools like Presidio Analyzer \cite{microsoft}and StarPII \cite{starpii} extend Named Entity Recognition (NER) capabilities by incorporating predefined dictionaries and regex patterns. However, such methods are inherently limited due to their dependence on static patterns, which fail to adapt to new contexts or nuanced PII definitions. There is also added complexity when a phone number may appear which may be identified correctly, but may not be considered as PII based on the context (e.g., phone number of an organization versus a private individual).

Context-aware PII detection has emerged as a significant advancement, leveraging machine learning techniques to incorporate the surrounding context of entities in text. Studies have explored BiLSTM-based models for analyzing forward and backward contexts, enabling more robust identification of PII in unstructured text \cite{patent1} as well as other methods \cite {gupta2021context, chen2023learning, yan2024ltner}. This contextual approach is essential for resolving ambiguities, such as distinguishing between a person’s name and a location when dealing with entities like "Paris" or "Jordan."

Despite advancements, existing tools still face critical gaps that limit their effectiveness in diverse real-world scenarios. One major issue is limited coverage across languages and jurisdictions. For example, a tool might perform well in detecting PII in English but struggle with languages like Arabic or Mandarin, and fail to adapt to regional regulations such as GDPR in Europe or CCPA in California. Another challenge lies in resolving conflicts between overlapping PII types. For example, a single string like \textit{"Jane Doe, 123-45-6789, Los Angeles"} might include a name, Social Security Number (SSN), and location, but tools often prioritize one type over others, resulting in incomplete or incorrect masking. Additionally, sensitivity scoring mechanisms for classifying data based on contextual importance remain inadequate. As highlighted in patents such as \cite{patent2} and \cite{patent3}, tools often lack the ability to dynamically adjust the sensitivity of data. For instance, they might treat an email address in a marketing email the same as an email address in confidential corporate communications, ignoring the vastly different contextual risks. 

\textbf{Privacy Risks in LLM Prompts and Outputs:} Privacy challenges in LLMs extend beyond entity recognition to include vulnerabilities such as prompt injection attacks, where malicious inputs exploit models to reveal sensitive information. Research on Propile \cite{kim2024propile}, a tool for probing privacy leakage in LLMs, highlights the risks of indirect prompt injections that can compromise user trust. Similarly, studies have underscored the challenges of adhering to privacy regulations like GDPR and CCPA, especially when dealing with multilingual data or region-specific compliance requirements. 

\textbf{Previous Deployments and Real-World Case Studies:} Deploying privacy-preserving frameworks in real-world scenarios remains underexplored, with limited studies addressing practical challenges. Our deployments represent two notable deployments that showcase the application of privacy guardrails in enterprise and open-source environments. These deployments emphasize contextual scoring, policy-driven actions, and automated triaging to mitigate privacy risks.

Additionally, prior studies have focused on using classifiers like Logistic Regression and CNNs for entity detection and sensitivity analysis \cite{liu2021automated, microsoft}.
They integrate privacy-preserving mechanisms with data governance workflows, as seen in enterprise platforms like WatsonNLP \cite{watsonNLP}. There are studies that focus on feedback loops for improving detection accuracy, particularly in community-driven ecosystems like GitHub. 

Existing research on privacy-preserving techniques for LLMs reveals several critical gaps that hinder their applicability in real-world deployments. One significant limitation lies in addressing contextual ambiguities, such as accurately resolving conflicts between overlapping PII types, where entities like a sequence of digits could be interpreted as a credit card number or a simple numeric identifier depending on the context 
\cite{verma2023fusionmind, jin2023alignment}. Additionally, current tools also struggle to scale effectively across multilingual data and adapt to diverse regional regulations like GDPR in Europe or CCPA in California, making compliance in global contexts challenging.

Real-time responsiveness is another limitation, as many frameworks cannot detect and mitigate privacy risks within milliseconds, making them impractical for enterprise-scale applications. Additionally, the lack of standardized benchmarks for evaluating tools across languages, domains, and PII types complicates the validation and comparison of their performance, leaving gaps in ensuring reliability and effectivenes \cite{edemacu2024privacy}. 

To address these gaps, this paper examines the OneShield Privacy Guard framework through two distinct deployments. By comparing their architectures, privacy-preservation techniques, and real-world impact, we aim to provide actionable insights for advancing privacy-preserving AI systems. Our work contributes to filling the gap in practical, deployment-focused studies that integrate technical, ethical, and compliance considerations.

%% file: deployment1.tex
\section{Deployment 1: \DMF}
\label{sec:deployment1}
%\vspace{-1mm}

\begin{figure*}
 \centerline{\includegraphics[scale=0.5]{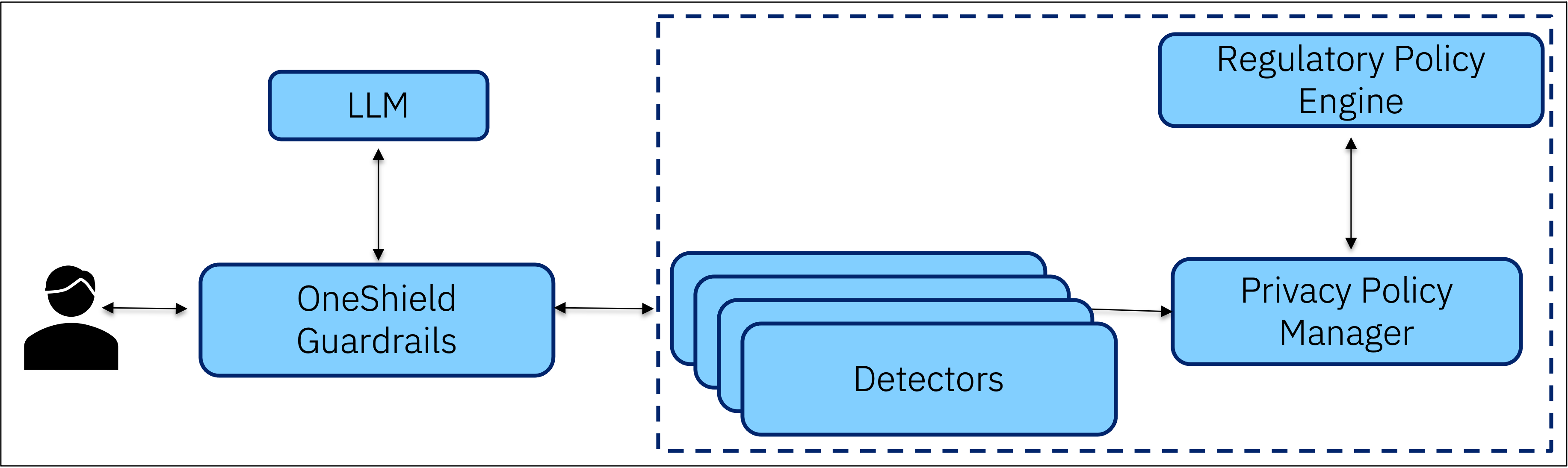}}
\caption{System Architecture for Deployment 1}
\label{fig:sysarch}
\end{figure*}

\begin{figure}
 \centerline{\includegraphics[scale=0.45]{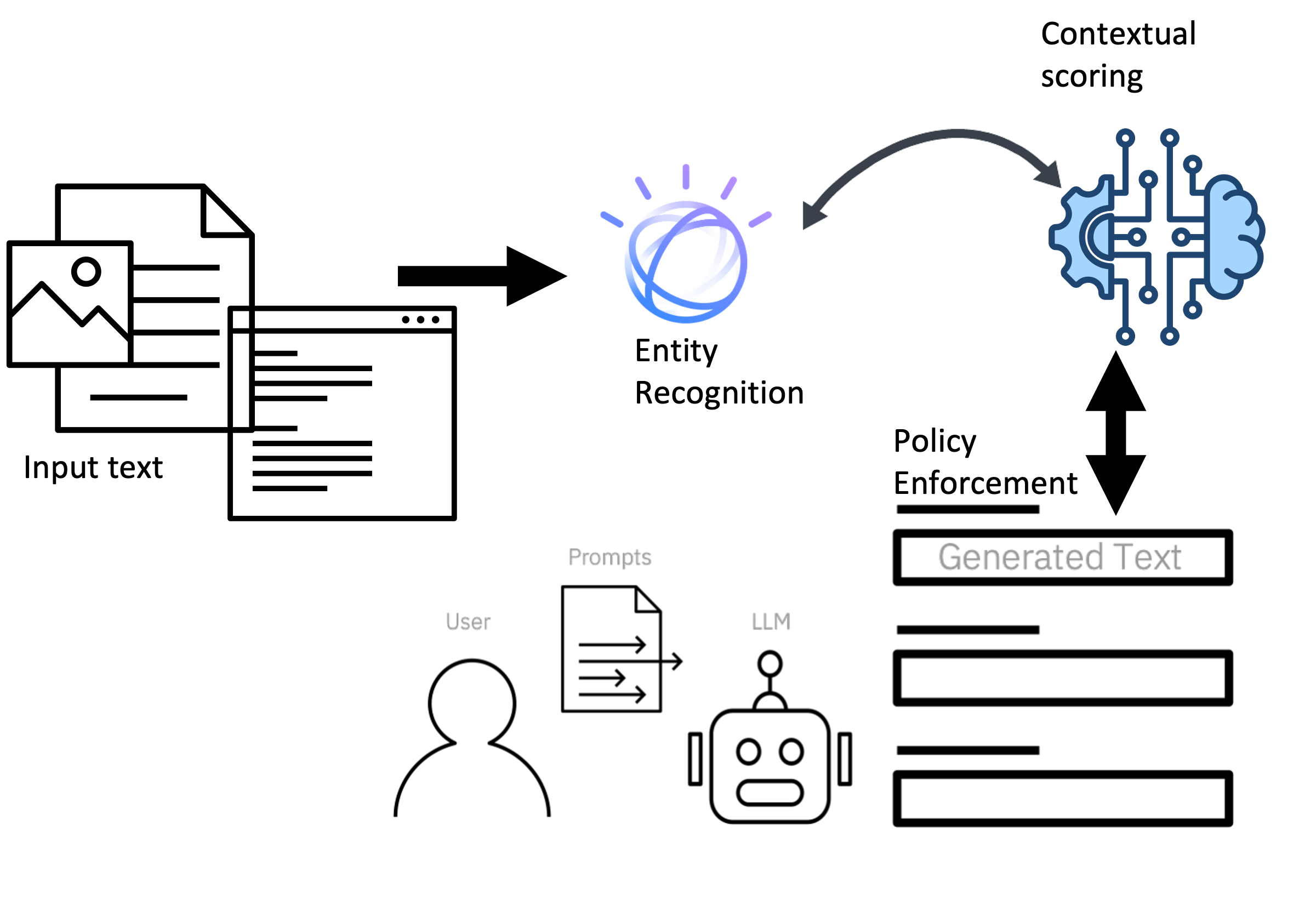}}
\caption{PII Detection pipeline}
\label{fig:piidetection}
\end{figure}

The OneShield Privacy Guard framework was first deployed within the \DMF, an internal platform supporting the creation and management of AI models and datasets with strong data governance measures. This platform facilitates the handling of diverse datasets, models, and LLM chatbot interactions while offering essential features like transparency, metadata tracking, and risk analysis across AI workflows. Given the platform’s heavy reliance on unstructured textual data, safeguarding PII emerged as a critical priority.

%The integration of OneShield addressed this need by providing a robust guardrail solution to protect PII within both user inputs and model outputs. This deployment played a vital role in supporting enterprise use cases enabled by the platform, such as customer service automation, personalized recommendations, and compliance monitoring tools. These applications frequently process sensitive data, making effective PII protection essential for adhering to stringent privacy regulations and mitigating risks. By focusing on entity recognition for specific types of PII, OneShield ensured precise, context-aware protection tailored to the unique requirements of these use cases, reinforcing the platform’s commitment to responsible AI.

The integration of OneShield provided a robust guardrail solution to protect PII in both user inputs and model outputs, addressing the diverse needs of enterprise-scale applications. This deployment supported use cases such as customer service automation, personalized recommendations, compliance monitoring, and AI-driven insights. The platform processes a variety of sensitive data types, including legal documents, books, web-crawled content, patents, financial records, and code, across 13 languages such as Arabic, Chinese, Czech, Dutch, English, Spanish, and Portuguese. These data sources are used for training and fine-tuning a range of models, including time series models, Large Language Models (LLMs), smaller LLMs, and specialized code models. Both base and fine-tuned versions of these models are hosted on the platform, enabling users to access data for training or retrieve models for processing tasks.

\subsection{System Architecture}
The OneShield deployment was built on a modular architecture that enabled flexibility and scalability. It consisted of three core components as shown in Figure \ref{fig:sysarch}:
\begin{enumerate}
    \item \textbf{Guardrail Solution}: Responsible for monitoring both input prompts and output responses to detect sensitive PII entities across multiple languages.
    \item \textbf{Detector Analysis Module}: Incorporated detection mechanisms for various PII types, along with additional capabilities like hate, abuse and profanity content detection, enabling a broader scope of privacy governance.
    \item \textbf{Privacy Policy Manager}: Provided policy templates tailored to jurisdictional regulations such as GDPR and CCPA, dynamically applying actions like masking, blocking, or passing input output data based on detected entities.
\end{enumerate}

This architecture supported seamless integration with the \DMF ecosystem, enabling it to process and protect data across $\sim$ 30 language models in real-time.

\subsection{Challenges in Deployment 1}
%Why we need this PII detection. What challenges were there that traditional approach could not do.  why this couldn't have been done using regex.. what was so unique that it required the OneShield framework technology.
Deployment 1 focused on protecting user inputs and model outputs across various AI use cases including patent data, books, web-crawled data etc. These applications often handle sensitive personal data like names, email addresses, and phone numbers. Detecting and masking this data was essential to comply with privacy regulations such as GDPR and CCPA while ensuring user trust.

Traditional methods like regex fell short in this scenario because they are rule-based and rigid. For example, a regex might detect \textit{"123-45-6789"} as a U.S. Social Security Number but fail to recognize variations in formatting or contexts, such as an SSN embedded within a larger text. Regex also struggles with multilingual data—detecting phone numbers in English is vastly different from identifying Indian Aadhaar IDs or German tax numbers. Moreover, overlapping data types, like a string containing both a name and a phone number, were beyond regex's capabilities to handle effectively.

The OneShield framework used contextual machine learning to overcome these challenges. Unlike regex, it could analyze the context to distinguish between sensitive data and similar-looking non-sensitive text. For instance, \textit{"John Smith"} could be flagged as a name only if used in a context suggesting personal identification, whereas regex might flag unrelated occurrences of \textit{"Smith"}. Its ability to support 26 languages and resolve complex overlaps made it uniquely suited for this deployment.

\subsection{PII Detection}
In the Detectors Analysis module, the PII detection is an important component with an iterative algorithm to detect sensitive information. The key steps as shown in Figure \ref{fig:piidetection} included:

\begin{itemize}
    \item \textbf{Entity Recognition:} Entities were detected in user prompts and model outputs using a combination of rule-based and machine learning classifiers. The system was capable of identifying a wide range of PII types, including names, dates, locations, and alphanumeric identifiers like bank account and credit card numbers.
    \item \textbf{Contextual Scoring:} Detected entities were evaluated for sensitivity using context-aware techniques, which ensured that the relationships between entities (e.g., names and dates of birth) were accurately captured to determine their privacy risk.
    \item \textbf{Policy Enforcement:} The Privacy Policy Manager as shown in Figure \ref{fig:sysarch} dynamically applied regulatory rules to detected entities. For instance, data deemed highly sensitive under GDPR like person along with his bank account number, or person with his national number etc. was masked before being stored or used in model inference.
\end{itemize}

\begin{figure}
 \centerline{\includegraphics[scale=0.6]{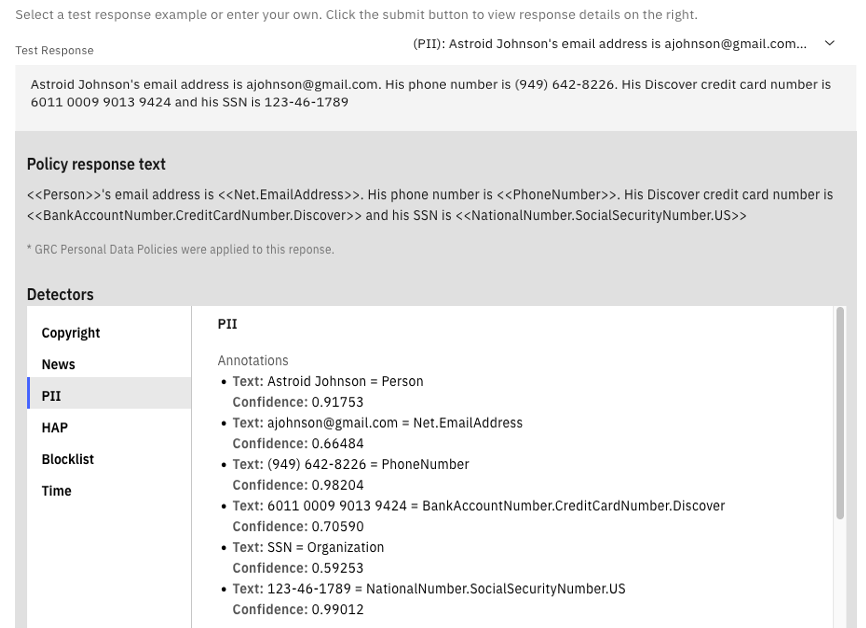}}
\caption{OneShield Guardrails deployment - PII detection}
\label{fig:flow}
\end{figure}

\subsection{Results and Evaluation}
The OneShield deployment demonstrated its effectiveness in safeguarding sensitive data through robust evaluation measures, including human assessments. Here, effectiveness refers to the system's ability to accurately detect and protect PII across various contexts, languages, and data formats, ensuring compliance with privacy standards while maintaining performance.  Figure \ref{fig:flow} gives a snapshot of the PII detection. The framework achieved high accuracy, consistently exceeding a 95\% detection rate across 26 languages, including English, Spanish, French, Turkish, Hindi, and Arabic. This performance highlights its capability to handle diverse and complex scenarios beyond the limitations of simpler solutions like regex.

On a dataset of $\sim$1,200 user prompts, the average response time for PII detection in inputs of up to 150 tokens was \textit{0.521} milliseconds, increasing to \textit{0.711} milliseconds for longer prompts of up to 250 tokens. The minimal increase in average responses is still within the non-functional requirements for real-time processing, adding negligible latency compared to baseline system performance without PII detection. Specifically, the additional latency was less than 5\% of the overall response time, ensuring a seamless user experience without compromising the system’s usability in high-throughput environments.

Additionally, the system provided detailed templates for masking PII and tracking policy violations, contributing to improved compliance management. Compared to open-source tools like Presidio Analyzer\cite{microsoft}, the OneShield Privacy Guard achieved superior coverage, particularly for complex PII types like national IDs and phone numbers from countries. For e.g., countries like Germany's phone number regex definition overlaps with India's Aadhar card definition, which were resolved using context. 

This deployment showed how important it is to support multiple languages and use context to improve privacy protection. By including policies tailored to specific regions, it used human evaluation to ensure the system met global privacy laws. The flexible design also made it easy to scale and use in different enterprise environments.

%% file: deployment2.tex
\section{Deployment 2: \PR Deployment}
\label{sec:deployment2}
%\vspace{-1mm}

The second deployment of the OneShield Privacy Guard framework was implemented in collaboration with \textit{\InstructLab}, an open-source repository hosted on GitHub. \textit{\InstructLab} provides a platform for community-driven contributions, including training datasets and seed examples for large language models. These contributions often take the form of pull requests (PRs) containing contexts, questions, and answers submitted by users. Due to the open nature of these contributions, there was a significant risk of sensitive personal information being inadvertently included, leading to potential violations of privacy or project codes of conduct.

To address these challenges, the OneShield guardrails solution was employed as a safety bot to detect and mitigate privacy risks in PRs. This deployment emphasized automating the detection of PII within the actual text of PR, and not metadata of PR, while reducing the burden on human triage teams.

\subsection{System Architecture}
The architecture of this deployment was tailored to handle community-generated content and featured two primary components:
\begin{enumerate}
    \item \textbf{Automated Detection Bot:} Integrated with the GitHub repository to scan incoming PRs in real time. The bot employed rule-based and machine-learning models to identify sensitive information within textual content.
    \item \textbf{Human-in-the-Loop Feedback Mechanism:} Detected PII were flagged for review by a triage team. This iterative approach allowed for fine-tuning the detection models based on human feedback and contextual relevance
\end{enumerate}

Unlike Deployment 1, this system prioritized lightweight processing to accommodate the high volume of PRs while maintaining compatibility with existing repository workflows.

\subsection{Challenges in Deployment 2}
Code repositories often included sensitive information like phone numbers, email addresses, or placeholders (e.g., \textit{"Albert Einstein's phone number: 123-456-7890"}). With hundreds of PRs submitted globally, automating privacy risk detection became essential to streamline reviews and ensure compliance.

Regex struggled to differentiate real sensitive data from irrelevant or fictional text. For example, regex might flag \textit{"Sherlock Holmes, 221B Baker Street"} incorrectly or miss actual violations like \textit{"Winston Churchill's phone: 987-654-3210"} hidden in comments. Overlapping formats, such as dates doubling as IDs (e.g., \textit{"July 4, 1776"}), added to its limitations.

OneShield used context to distinguish genuine risks from false positives, flagging \textit{"Mahatma Gandhi's phone: +91-1234567890"} as sensitive while ignoring placeholders like \textit{"Jane Austen's email: jane.austen@example.com."} Its multilingual support and precision reduced false positives, saved over 300 hours of manual reviews, and ensured compliance across global teams.

\subsection{PII detection}
The PII detection process involved three key steps:

\begin{itemize}
    %\item \textbf{Entity Recognition:} The system parsed text into smaller chunks using separators like periods and line breaks, enabling sentence-level entity detection. The entities were categorized into PII types such as names, email addresses, phone numbers, and dates. 
    \item \textbf{Entity Recognition:} The system broke text into smaller chunks using separators like periods and line breaks to detect entities like names, email addresses, phone numbers, and dates. Monitoring phone numbers and email addresses is important because developers may accidentally include them in PRs, such as using real or placeholder contact details in examples or test cases (e.g., \textit{"John Doe, john.doe@example.com, 123-456-7890"}). Other sensitive PII in PRs includes names, dates, and confidential keys like API tokens. Detecting and flagging this information helps prevent privacy breaches and ensures compliance with project guidelines.
    \item \textbf{Contextual Sensitivity Scoring:} Detected entities were evaluated for sensitivity based on their relationships with other text elements. For example, \textit{"John Doe"} combined with an email address was classified as highly sensitive, while standalone generic entities like \textit{"Date : 1776"} were deemed non-sensitive.
    \item \textbf{Automated Policy Actions:} Detected violations were flagged, preventing PRs containing sensitive information from being merged into the repository until reviewed. This ensured compliance with the project’s code of conduct, which prohibited the inclusion of private or invasive content.
\end{itemize}

\begin{table*}[htb]
\footnotesize
    \centering
\begin{tabular}{p{0.2\textwidth} p{0.18\textwidth} p{0.18\textwidth} p{0.15\textwidth} p{0.15\textwidth}} 
    \hline
    \textbf{Text} & \textbf{Detected Entity} & \textbf{Sensitivity Level} & \textbf{Policy Compliance} & \textbf{Action Taken}\\
    \hline
    "Alice visited Paris on January 12, 2023." & Name: Alice & Level 2 (Sensitive) & Policy Violation Detected & Masked Name \\
    "Contact me at john.doe@example.com for details." & Email Address: john.doe & Level 1 (Highly Sensitive) & Policy Violation Detected & Email Address Masked \\
    "Barack Obama was born on August 4, 1961." & Name: Barack Obama & Level 3 (Non-Sensitive, Public Figure) & Policy Compliant & No Action Taken\\
    "The account balance as of March 5 was \$12,000." & Date: March 5 & Level 2 (Sensitive) & Policy Violation Detected & Date Masked\\
    "Historical events from 1776 included the Declaration of Independence." & Date: 1776 & Level 3 (Non-Sensitive) & Policy Compliant & No Action Taken\\
    \bottomrule
\end{tabular}
\caption{ Detected PII entities in PR Insights Deployment}
\label{tab:table1}
\end{table*}

\subsection{Results and Evaluation}

In Deployment 2, the system analyzed over 1,256 pull requests (PRs) within a span of three months, identifying privacy violations in 8.25\% of cases. The most commonly flagged entities included names, phone numbers, and dates. By automating the initial detection of potential privacy violations, the deployment significantly reduced the workload of the triage team, pre-flagging PRs for further review and action.

This automation saved an estimated 300+ hours of manual effort over the three-month period, assuming that manually reviewing a PR for privacy violations typically takes 15 minutes. The streamlined process allowed the team to focus on higher-priority tasks, improving overall efficiency. Furthermore, the proactive identification of privacy risks prevented any  potential breaches, reducing the likelihood of costly penalties or reputational damages. In terms of business impact, this deployment enhanced operational productivity by approximately 25\%, ensuring quicker turnaround times for PR approvals and fostering a culture of privacy-by-design in development workflows.

Table \ref{tab:table1} showcases detected PII entities, their contextual sensitivity scores, and compliance with defined policies. It includes anonymized examples from Deployment 2, illustrating how the system flagged violations and took appropriate actions like masking or blocking sensitive data. The system flagged "Alice visited Paris on January 12, 2023," identified the name and date as Level 2 sensitive due to their contextual combination. Similarly, "Contact me at john.doe@example.com" was flagged as a policy violation and the email was masked to prevent unauthorized disclosure.

Feedback loops from the triage team played a critical role in refining the detection models. For instance, early versions of the system misclassified historical figures and fictional characters as sensitive entities. Iterative updates to the model resolved these inaccuracies by incorporating contextual cues and public knowledge databases like Wikimedia SPARQL APIs.

This deployment underscored the importance of human-in-the-loop systems in refining automated PII detection. By iteratively improving classifier models and contextual analysis, the framework achieved better precision in distinguishing sensitive information from non-sensitive content. Furthermore, embedding privacy safeguards into community-driven platforms highlighted the value of automated tools in maintaining ethical standards while supporting open-source collaboration.

%% file: comparison.tex
\section{Comparison of Deployments and Discussion}
\label{sec:comparison}
%\vspace{-1mm}

\subsection{Comparison of Deployments}
The two deployments of the OneShield Privacy Guard framework showcase different approaches tailored to their respective application environments. The first deployment (\DMF) emphasized enterprise-scale governance and multilingual adaptability, while the second (\PR) focused on automating privacy checks in open-source contributions.

\begin{table*}[htb]
\footnotesize
    \centering
\begin{tabular}{p{0.15\textwidth} p{0.2\textwidth} p{0.2\textwidth} p{0.1\textwidth} p{0.15\textwidth}} 
    \toprule
    \textbf{PII Type} & \textbf{LLM Privacy Guard (D1)} & \textbf{LLM Privacy Guard (D2)} & \textbf{StarPII} & \textbf{Presidio Analyzer}\\
    \midrule
     Person & 0.98 & 0.95 & 0.87 & 0.91\\
     Date & 0.96 & 0.88 & 0.85 & 0.72 \\
     Email Address & 0.94 & 0.89 & 0.73 &0.85 \\
     Phone Number & 0.89 & 0.82 & 0.71 & 0.77 \\
     Location & 0.91 & 0.84 & 0.74 & 0.81 \\
     National ID & 0.92 & 0.78 & N/A & N/A\\
     Credit Card & 0.95 & N/A & 0.68 & 0.72\\
    \bottomrule
\end{tabular}
\caption{Comparison of PII detection F1 scores with existing state-of-the-art PII detectors}
\label{tab: table2}
\end{table*}

We performed a comparison of our PII detection in the two deployments with two state-of-the-art PII detectors, StarPII \cite{starpii} and Presidio Analyzer\cite{microsoft}, across various PII types. A high-level summary of results are seen in Table \ref{tab: table2}. The data showcases the higher accuracy of OneShield Privacy Guard for both Deployment 1 and 2, as compared to StarPII and Presidio Analyzed. The performance was observed for sensitive entities, particularly in multilingual and regulatory-complex contexts. The results have been updated with deployment-specific benchmarks from Deployment 1 (\DMF) and Deployment 2 (\PR). In Deployment 1, the OneShield Privacy Guard demonstrated a 0.95 F1  score in detecting dates across multilingual data, such as identifying sensitive dates in Spanish and French prompts. Deployment 2 showed overall better performance in detecting email addresses, phone numbers, persons, location etc. within open-source PRs, where contributors often accidentally included personal contact details.

\begin{table*}[htb]
\footnotesize
    \centering
\begin{tabular}{p{0.15\textwidth} p{0.38\textwidth} p{0.38\textwidth}} 
    \toprule
    \textbf{Metric} & \textbf{Deployment 1: \DMF} & \textbf{Deployment 2: \PR} \\
    \midrule
    Objective & Safeguard PII in enterprise-scale multilingual outputs.& Ensure privacy compliance in open-source GitHub pull requests. \\
    Scope & 30 language models; multilingual data handling. & Community-generated unstructured text in English. \\
    Challenges & Managing diverse regulatory requirements and ensuring consistent accuracy across languages.& Handling ambiguities in public vs. private information and improving classification of sensitive data.\\
    System Architecture & Modular design with a Privacy Policy Manager and multilingual support. & Lightweight bot integrated with a GitHub repository workflow. \\
    PII Detection & High accuracy in detecting context-sensitive PII across languages. & Focused detection of PII in PRs; iterative refinement through feedback.\\
    Latency & 0.521 ms for 150 tokens; 0.711 ms for 250 tokens. & Processed over 1,256 PRs; flagged 8.25\% for potential violations.\\
    Policy Enforcement & Dynamic enforcement of GDPR, CCPA, and PIPEDA compliance. & Automated triaging with human-in-the-loop validation for code of conduct adherence.\\
    Feedback Loop & Minimal human feedback due to policy-driven automation.& Extensive feedback from human reviewers for model refinement.\\
    Key Outcomes & Enhanced scalability for enterprise applications; robust multilingual PII coverage.& Streamlined privacy compliance in collaborative environments; reduced manual effort.\\
    \bottomrule
\end{tabular}
\caption{Comparative analysis of the two deployments}
\label{tab: table3}
\end{table*}

\subsection{Discussion}
We also performed a comparison of overall deployments. Table \ref{tab: table3} provides discussion on the comparative analysis. The comparative analysis reveals critical insights into the challenges and opportunities of deploying privacy-preserving frameworks in varied environments:
\begin{enumerate}
    \item \textbf{Scalability and Multilingual Support:} Deployment 1 demonstrated the scalability of OneShield Privacy Guard across languages and regulatory landscapes, providing a robust solution for enterprise needs. However, it faced challenges in maintaining consistent accuracy across diverse data types. Deployment 2, while monolingual, highlighted the effectiveness of lightweight systems in collaborative workflows, particularly when combined with human-in-the-loop processes for model refinement.
    \item \textbf{Compliance and Regulatory Adherence:} Both deployments underscored the importance of aligning privacy-preserving tools with region-specific regulations. Deployment 1 leveraged a Privacy Policy Manager to dynamically enforce compliance, while Deployment 2 provided automated triaging that helped maintain the ethical standards of community-driven platforms like GitHub.
    \item \textbf{Lessons in Contextual Sensitivity:} Contextual sensitivity emerged as a key factor in both deployments. For instance, Deployment 1 resolved ambiguities by analyzing relationships between entities (e.g., combining names with date of birth). Deployment 2 faced challenges in classifying historical and fictional entities, which were resolved through iterative updates informed by human feedback.
    \item \textbf{Automation vs. Human Oversight:} Automation was a cornerstone of both deployments, but human oversight proved critical in Deployment 2 for refining classifiers and addressing edge cases. Eg Pseuanoanonyms are protected under GDPR, while not covered in CCPA. However this can get missed in automation. This highlights the need for hybrid approaches that balance automation with human validation.
    \item \textbf{Future Directions for Privacy-Preserving AI:} The insights from these deployments point to the necessity of extending privacy-preserving frameworks to handle multimodal data, such as images and audio. Additionally, integrating adaptive, self-learning systems to address evolving privacy laws and cultural sensitivities will be crucial for future advancements.
\end{enumerate}

%% file: conclusion.tex
\section{Conclusions and Future Work}
\label{sec:conclusion}

This paper presents a comparative analysis of two real-world deployments of the OneShield Privacy Guard framework, emphasizing their role in preserving privacy across diverse operational contexts. Deployment 1, integrated into \DMF, demonstrated the scalability and adaptability required for enterprise-level applications, particularly in multilingual and regulatory-complex environments. Deployment 2, implemented in \PR, highlighted the effectiveness of lightweight privacy solutions in community-driven platforms, leveraging human feedback to refine automated PII detection.

Both deployments highlight the importance of using context-aware entity recognition, dynamic policy enforcement, and a balance between automation and human oversight. They address key challenges like ambiguity in sensitive data classification and compliance with regional regulations, offering valuable insights for building privacy-preserving frameworks in LLMs.

Future advancements should focus on expanding privacy safeguards to handle multimodal data, including images, audio, and video, to address risks in cross-modal AI systems like vision-language models. Self-learning mechanisms, such as reinforcement learning, can help frameworks adapt to changing regulations and cultural sensitivities, reducing manual intervention while maintaining accuracy. Finally, creating standardized benchmarks with multilingual and cross-domain datasets will improve the evaluation and comparability of privacy-preserving tools. These steps are essential for developing scalable, ethical, and effective AI technologies in complex data environments.

\section{Acknowledgements}
\label{sec:ack}
We would like to thank Guang-Jie Ren and Pawan Chowdhary for their guidance and contributions on this work.

%% file: aaai25.bbl
\begin{thebibliography}{22}
\providecommand{\natexlab}[1]{#1}

\bibitem[{wat(2022)}]{watsonNLP}
 2022.
\newblock IBM WatsonNLP.
\newblock
  \url{https://developer.ibm.com/tutorials/entity-extraction-using-watson-nlp/}.

\bibitem[{com(2023)}]{comprehend}
 2023.
\newblock AWS Comprehend.
\newblock \url{https://docs.aws.amazon.com/comprehend/latest/dg/how-pii.html}.

\bibitem[{sta(2023)}]{starpii}
 2023.
\newblock HuggingFace StarPII.
\newblock \url{https://huggingface.co/bigcode/starpii}.
\newblock Microsoft Presidio.

\bibitem[{neu(2023)}]{neuralseek}
 2023.
\newblock NeuralSeek PII detection.
\newblock
  \url{https://documentation.neuralseek.com/main_features/advanced_features/advanced_features/#pii-detection
  }.

\bibitem[{mic(2023)}]{microsoft}
 2023.
\newblock RPii detection cognitive skill - azure cognitive search, 2022a.
\newblock \url{https://learn.microsoft.com/
  en-us/azure/search/cognitive-search-skill-pii-detection}.
\newblock Microsoft Presidio.

\bibitem[{ccp(2024)}]{ccpa}
 2024.
\newblock California Consumer Privacy Act.
\newblock \url{https://oag.ca.gov/privacy/ccpa}.
\newblock California Consumer Privacy Act.

\bibitem[{gdp(2024)}]{gdpr}
 2024.
\newblock General Data Protection Regulation.
\newblock \url{https://gdpr-info.eu/}.
\newblock General Data Protection Regulation.

\bibitem[{pip(2024)}]{pipeda}
 2024.
\newblock The Personal Information Protection and Electronic Documents Act
  (PIPEDA).
\newblock
  \url{https://www.priv.gc.ca/en/privacy-topics/privacy-laws-in-canada/the-personal-information-protection-and-electronic-documents-act-pipeda/}.
\newblock The Personal Information Protection and Electronic Documents Act
  (PIPEDA).

\bibitem[{Carlini et~al.(2022)Carlini, Ippolito, Jagielski, Lee, Tramer, and
  Zhang}]{carlini2022quantifying}
Carlini, N.; Ippolito, D.; Jagielski, M.; Lee, K.; Tramer, F.; and Zhang, C.
  2022.
\newblock Quantifying memorization across neural language models.
\newblock \emph{arXiv preprint arXiv:2202.07646}.

\bibitem[{Carlini et~al.(2021)Carlini, Tramer, Wallace, Jagielski,
  Herbert-Voss, Lee, Roberts, Brown, Song, Erlingsson
  et~al.}]{carlini2021extracting}
Carlini, N.; Tramer, F.; Wallace, E.; Jagielski, M.; Herbert-Voss, A.; Lee, K.;
  Roberts, A.; Brown, T.; Song, D.; Erlingsson, U.; et~al. 2021.
\newblock Extracting training data from large language models.
\newblock In \emph{30th USENIX Security Symposium (USENIX Security 21)},
  2633--2650.

\bibitem[{Chen et~al.(2023)Chen, Lu, Lin, Lou, Jia, Dai, Wu, Cao, Han, and
  Sun}]{chen2023learning}
Chen, J.; Lu, Y.; Lin, H.; Lou, J.; Jia, W.; Dai, D.; Wu, H.; Cao, B.; Han, X.;
  and Sun, L. 2023.
\newblock Learning in-context learning for named entity recognition.
\newblock \emph{arXiv preprint arXiv:2305.11038}.

\bibitem[{Edemacu and Wu(2024)}]{edemacu2024privacy}
Edemacu, K.; and Wu, X. 2024.
\newblock Privacy preserving prompt engineering: A survey.
\newblock \emph{arXiv preprint arXiv:2404.06001}.

\bibitem[{et~al.(US10984316B2, 2017)}]{patent3}
et~al., M.~Q. US10984316B2, 2017.
\newblock Context aware sensitive information detection.

\bibitem[{et~al.(US11127403B2, 2019)}]{patent1}
et~al., S.~M. US11127403B2, 2019.
\newblock Machine learning-based automatic detection and removal of personally
  identifiable information.

\bibitem[{Gupta et~al.(2021)Gupta, Verma, Mashetty, and
  Mishra}]{gupta2021context}
Gupta, H.; Verma, S.; Mashetty, S.; and Mishra, S. 2021.
\newblock Context-ner: Contextual phrase generation at scale.
\newblock \emph{arXiv preprint arXiv:2109.08079}.

\bibitem[{Jin et~al.(2023)Jin, Cao, He, Chen, Liu, and Zhao}]{jin2023alignment}
Jin, Z.; Cao, P.; He, Z.; Chen, Y.; Liu, K.; and Zhao, J. 2023.
\newblock Alignment Precedes Fusion: Open-Vocabulary Named Entity Recognition
  as Context-Type Semantic Matching.
\newblock In \emph{Findings of the Association for Computational Linguistics:
  EMNLP 2023}, 14616--14637.

\bibitem[{Kim et~al.(2024)Kim, Yun, Lee, Gubri, Yoon, and Oh}]{kim2024propile}
Kim, S.; Yun, S.; Lee, H.; Gubri, M.; Yoon, S.; and Oh, S.~J. 2024.
\newblock Propile: Probing privacy leakage in large language models.
\newblock \emph{Advances in Neural Information Processing Systems}, 36.

\bibitem[{Liu et~al.(2021)Liu, Lin, Ebrahimi, Li, and Chen}]{liu2021automated}
Liu, Y.; Lin, F.~Y.; Ebrahimi, M.; Li, W.; and Chen, H. 2021.
\newblock Automated pii extraction from social media for raising privacy
  awareness: A deep transfer learning approach.
\newblock In \emph{2021 IEEE International Conference on Intelligence and
  Security Informatics (ISI)}, 1--6. IEEE.

\bibitem[{Ron M.~Redlich(US8468244B2, 2009)}]{patent2}
Ron M.~Redlich, M. A.~N. US8468244B2, 2009.
\newblock Digital information infrastructure and method for security designated
  data and with granular data stores.

\bibitem[{Smith et~al.(2013)Smith, Saint-Amand, Plamada, Koehn, Callison-Burch,
  and Lopez}]{smith2013dirt}
Smith, J.~R.; Saint-Amand, H.; Plamada, M.; Koehn, P.; Callison-Burch, C.; and
  Lopez, A. 2013.
\newblock Dirt cheap web-scale parallel text from the common crawl.
\newblock Association for Computational Linguistics.

\bibitem[{Verma et~al.(2023)Verma, Parmar, Choudhary, and
  Porwal}]{verma2023fusionmind}
Verma, S.; Parmar, M.; Choudhary, P.; and Porwal, S. 2023.
\newblock FusionMind--Improving question and answering with external context
  fusion.
\newblock \emph{arXiv preprint arXiv:2401.00388}.

\bibitem[{Yan, Yu, and Chen(2024)}]{yan2024ltner}
Yan, F.; Yu, P.; and Chen, X. 2024.
\newblock LTNER: Large Language Model Tagging for Named Entity Recognition with
  Contextualized Entity Marking.
\newblock \emph{arXiv preprint arXiv:2404.05624}.

\end{thebibliography}
